\documentclass[12pt]{article} 
\usepackage{latexsym,cite,subeqn}
\input amssym.def    
\input amssym.tex  
\makeatletter   
\renewcommand{\thesection}{\Roman{section}.}

\renewcommand{\theequation}{\thesection\arabic{equation}} 
\@addtoreset{equation}{section} 
\makeatother  
\font\msym=msbm10

\def\Integer{{\mathop{\hbox{\msym \char  '132}}}}

\def\L{{\cal L}}

\def\pr{{\textstyle{\frac{2\pi}{R}}}}

\begin{document}  
\begin{titlepage}
\title{\vskip -60pt
{\small\begin{flushright} 
KIAS-P00054\\
hep-th/0008103 
\end{flushright}}
\vskip 45pt
$5$\,D Actions for $6$\,D Self-Dual Tensor Field Theory\\
~\\}
\vspace{4.0cm}
\author{\\
\\
\\Kimyeong Lee and Jeong-Hyuck Park}
\date{}
\maketitle
\vspace{-1.0cm}
\begin{center}
\textit{Korea Institute for Advanced Study}\\
\textit{207-43 Cheongryangri-dong Dongdaemun-gu, Seoul 130-012, Korea}
\end{center}
\vspace{2.0cm}
\begin{abstract}
\noindent We present two  equivalent  five-dimensional actions for six-dimensional  $(N,0)$~$N=1,2$ supersymmetric theories of self-dual tensor whose one spatial dimension is compactified on a circle.   The Kaluza-Klein tower consists of a massless vector and infinite number of massive self-dual tensor multiplets living in five-dimensions.  
%%%
%%The actions are written in five-dimensions in terms of the
%%% Kaluza-Klein modes of the  tensor-multiplets. 
%%%%
The self-duality follows from the equation of motion.  
Both  actions are quadratic in field variables without any auxiliary field. When lifted back to six-dimensions, one of them  gives  a  supersymmetric extension of the bosonic formulation for the chiral two-form tensor by Perry and Schwarz. 
%%%%% 
%%We argue that there is a hidden big symmetry in 
%%$6$\,D theory which contains both the superconformal
%%symmetry and some local symmetries. 
%%%%
\end{abstract}
\thispagestyle{empty}
\end{titlepage}
\newpage

\section{Introduction and Conclusion}
One of the challenging problems in quantum field theories at present  is  to construct  the action for chiral $p$-forms, i.e. anti-symmetric boson fields whose  field strength is self-dual, especially with non-Abelian group structure implemented. Self-duality condition requires the spacetime to be Euclidean for odd $p$ and Minkowskian for even $p$.  In particular, $p=2$ case has been of much attention related to the formulation of the world-volume action for M-theory five-brane~\cite{PLB27649,NPB36760,PRL713754,9510053,9602071,9610234,9611008}.\newline

\indent Concerning Abelian theories of chiral $p$-forms there have been various types of proposals.  Floreanini and Jackiw first proposed a non-manifestly covariant action for chiral scalars in two-dimensions by adopting somewhat unusual commutation relations  among the field variables~\cite{PRL591873}.   McClain, Wu and Yu proposed a formulation for chiral scalars  by  introducing an infinite number of auxiliary   fields which do not carry any physical degree of freedom~\cite{NPB343689}.  Each treatment was extended to higher order $p$-forms in Refs.~\cite{PLB206650} and \cite{9603031,9609102} respectively.       \newline

\indent Two other  formulations are also available. Pasti, Sorokin and Tonin introduced a Lorentz covariant formulation  with only one auxiliary scalar field entering a chiral $p$-form action in a non-polynomial way~\cite{9611100}.  Schwarz \textit{et al.} studied a non-covariant formulation for self-dual two-form tensor in six-dimensions, where only five-dimensional Lorentz symmetry is manifest as one spacetime dimension is treated differently from the others~\cite{9304154,9611065,9701008}. However, it turned out  that the PST formulation for $p=2$ case contains local symmetries  and a non-covariant gauge fixing of the local symmetries reduces to the non-manifestly Lorentz invariant  formulation by Schwarz \textit{et al.}~\cite{9701037,9701166}(see also \cite{9503182,9506109,9509052}).  Each formulation further developed  to construct a kappa symmetric  world-volume action for M-theory five-brane in an eleven-dimensional  superspace background~\cite{9701149,9701166}.  \newline

\indent The bosonic PST formulation was  supersymmetrized in six-dimensions incorporating the self-dual  tensor-multiplets.  Dall'Agata \textit{et al.} and Claus \textit{et al.} presented  the $(1,0)$ and $(2,0)$  supersymmetric extensions separately~\cite{9710127},~\cite{9711161}. On the other hand,  the  non-manifestly Lorentz invariant action by Schwarz \textit{et al.} has not been supersymmetrized in the literature yet.  \newline

\indent In the context of M-theory, five-dimensional maximally   supersymmetric  gauge theory at strong coupling limit is supposed to have description by a six-dimensional $(2,0)$ fixed point, as the four-brane of type IIA theory is the M-theory five-brane wrapped around the eleventh direction and at strong coupling the eleventh dimension decompactifies developing an extra dimension~\cite{9702136,9704089}.  In fact, there is no interacting   fixed point of the renormalization group in five-dimensions~\cite{seiberg16}.  Nevertheless, direct  field theoretic understanding of the relationship between the five and six dimensional theories for non-Abelian interactions  is still lacking.   \newline

\indent In this paper, we present two different but equivalent  five-dimensional supersymmetric actions for the Kaluza-Klein modes of  the six-dimensional $(N,0)$, $N=1,2$ self-dual tensor-multiplets compactified on a circle.  The Kaluza-Klein tower consists of a massless vector and infinite number of massive self-dual tensor multiplets living in five-dimensions. The self-duality follows from the equation of motion rather than a constraint imposed by hand.  When lifted back to six-dimensions, one of our formulations   gives  a  supersymmetric extension of the bosonic action for the chiral two-form tensor by Perry and Schwarz~\cite{9611065}.  As there appears a five-dimensional vector multiplet after  compactifying  the $6$\,D  tensor-multiplet on a circle, one may try to implement the non-Abelian group structure by taking the vector field as the usual Yang-Mills gauge field.\footnote{In this approach, one needs to ensure the six-dimensional covariance of the non-Abelian gauge symmetry in the five-dimensional action.}  This would give a five-dimensional super Yang-Mills theory coupled with massive tensor-multiplets in an adjoint representation, realizing the M-theory picture on $5\,$D and $6\,$D theories.  This scenario is the main motivation of the work in the present paper. The proposed formulations deal with Abelian case. Supersymmetry is provided, and non-Abelian generalization is to be done. \newline

\indent In the following  section~\ref{compact}, we first compactify the $6$\,D  tensor-multiplets  on a circle. The self-duality is expressed in terms of the Kaluza-Klein modes  in five-dimensional language. The  resulting Kaluza-Klein modes are massless vector and massive tensor multiplets which are identified by the analysis on five-dimensional supersymmetry algebra.  In section \ref{sec10} and \ref{sec20} we write our two proposed actions for the Kaluza-Klein modes of the $(1,0)$ and $(2,0)$ tensor-multiplets  respectively.  In section \ref{lift} we lift the actions to six-dimensions and discuss the symmetries.

%%%%%%%%%%%%%%%%%%%%%%%%%%%%%%%%%%%%%%%%%%%%%%%%%%%%%%%%%%%%%%%%%%%%%%%%%%%%%%%%%%%%%%%%%%%%%%%%%%%%%%%%%%%%%%%%%%%%%%%%%%%%%%%%%%%%%%%%%%%%%%%%%%%%%%%%%%%%%%%%%%%%%%%%%%%%%%%%%%%%%%%%%%%%%%%%%%%%%%%%%%%%%%%%%%%%%%%%%%%%%%%%
\section{Tensor-multiplets Compactified on a Circle\label{compact}}
Using  the $4\times 4$  gamma matrices, 
$\gamma^{\mu},\,\mu=0,1,\cdots,4$, in five-dimensional Minkowskian spacetime with the metric, $\eta_{\mu\nu}=\mbox{diag}(+1,-1,\cdots,-1)$,   the  six-dimensional gamma matrices, $\Gamma^{\hat{\mu}},\,\hat{\mu}=\mu,5$, are taken here as
\begin{equation}
\begin{array}{ccc}
\Gamma^{\hat{\mu}}=\left(\begin{array}{ll}
0&\gamma^{\hat{\mu}}\\
\tilde{\gamma}^{\hat{\mu}}&0
\end{array}\right)\,,~~~~&~~~~\gamma^{\mu}=\tilde{\gamma}^{\mu}\,,~~~~\gamma^{5}=-\tilde{\gamma}^{5}=1\,.
\end{array}
\end{equation}
This choice of gamma matrices gives a diagonalized  $\Gamma^{7}$ matrix  so that the non-vanishing components of the  six-dimensional chiral spinors are upper four, $\psi$, only  and  the pseudo-Majorana \textit{or} symplectic $\mbox{sp}(N)$-Majorana condition for $6$\,D  $(N,0)$ chiral spinors is readily translated into the 
five-dimensional   pseudo-Majorana  condition~\cite{kugotownsend} 
\begin{equation}
\begin{array}{cc}
\bar{\psi}_{i}=\psi^{i}{}^{\dagger}\gamma^{0}=\psi^{j}{}^{t}CJ_{ji}\,,~~~~&~~~~
J_{ij}=\left(\begin{array}{cr}
0&-1\\
1&0
\end{array}\right)\,,
\end{array}
\label{PM}
\end{equation}
where $1\leq i,j\leq 2N$ and $C$ is the five-dimensional charge conjugate matrix satisfying $\gamma^{\mu}{}^{t}=C\gamma^{\mu}C^{-1}\,,C^{t}=-C$. \newline

$6\,$D $(N,0),~N=1,2$ tensor-multiplet consists of a two-form tensor, $B_{\hat{\mu}\hat{\nu}}$,  pseudo-Majorana chiral spinors, $\psi^{i}$, and  one  for $N=1$/five  for $N=2$ real scalar(s), $\phi$ \cite{NPB221331}.  \newline

\indent Compactifying the fifth spatial dimension on a circle of radius $R$ gives  
a Kaluza-Klein tower of the tensor-multiplets 
\begin{equation}
\begin{array}{ll}
B_{\hat{\mu}\hat{\nu}}=\displaystyle{\sum_{m\in\Integer}}B_{m\hat{\mu}\hat{\nu}}\,e^{i\pr mx^{5}}\,,~~~~&~~~~
\phi=\displaystyle{\sum_{m\in\Integer}}\phi_{m}\,e^{i\pr mx^{5}}\,,\\
{}&{}\\
\psi^{i}=\displaystyle{\sum_{m\in\Integer}}\psi^{i}_{m}\,e^{i\pr mx^{5}}\,,~~~~&~~~~
\bar{\psi}_{i}=\displaystyle{\sum_{m\in\Integer}}\bar{\psi}_{mi}\,e^{i\pr mx^{5}}\,.
\end{array}
\end{equation} 
Reality and Pseudo-Majorana conditions imply
\begin{equation}
\begin{array}{ccc}
B_{m\hat{\mu}\hat{\nu}}^{\,*}=B_{-m\hat{\mu}\hat{\nu}}\,,~~~&~~~
\phi_{m}^{\,*}=\phi_{-m}\,,~~~&~~~
\bar{\psi}_{mi}=\psi_{-m}^{i\dagger}\gamma^{0}=\psi_{m}^{j}{}^{t}CJ_{ji}\,.
\end{array}
\label{realpM}
\end{equation}
The self-duality of the $6$\,D two-form tensor, $H=* H$, is now expressed in terms of the five-dimensional Kaluza-Klein  modes
\begin{equation}
F_{m\mu\nu}+i\pr mB_{m\mu\nu}=\textstyle{\frac{1}{6}}\epsilon_{\mu\nu}{}^{\lambda\rho\sigma}H_{m\lambda\rho\sigma}\,,
\label{selfdual}
\end{equation}
where $F_{m\mu\nu}$ is the field strength of $B_{m\mu 5}\equiv A_{m\mu}$.  \newline

Taking a curl of eq.(\ref{selfdual}) eliminates $A_{m\mu}$ leaving a second order  differential equation that involves $B_{m\mu\nu}$ only
\begin{equation}
\partial_{\lambda}H_{m}^{\lambda\mu\nu}=i\textstyle{\frac{\pi}{3R}}m\,\epsilon^{\mu\nu}{}_{\lambda\rho\sigma}H_{m}^{\lambda\rho\sigma}\,.
\label{curl}
\end{equation}
Reversely,  taking off the curl,   eq.(\ref{curl}) implies  $\textstyle{\frac{1}{6}}\epsilon_{\mu\nu}{}^{\lambda\rho\sigma}H_{m\lambda\rho\sigma}-i\pr mB_{m\mu\nu}=F_{m\mu\nu}^{\prime}$ for a certain $F^{\prime}_{m\mu\nu}=\partial_{\mu}A_{m\nu}^{\prime}-\partial_{\nu}A_{m\mu}^{\prime}$. In $m\neq 0$ cases  one can fix the gauge for the two-form tensor such that  $F_{m\mu\nu}^{\prime}=F_{m\mu\nu}$, while  $m=0$ case in eqs.(\ref{selfdual},\ref{curl}) shows the hodge dual relation between the five-dimensional free Maxwell theory and a massless free two-form field theory. Thus, eq.(\ref{curl}) is equivalent to eq.(\ref{selfdual}) upto gauge transformations.\newline

Six-dimensional $(N,0)$ supersymmetry algebra naturally descends to five-dimensions
\begin{equation}
{}\{Q^{i},\bar{Q}_{j}\}=\delta^{i}{}_{j}\gamma^{\hat{\mu}}P_{\hat{\mu}}=\delta^{i}{}_{j}(\gamma^{\mu}P_{\mu}+M)\,,
\end{equation}
where the supercharges, $Q^{i},\,1\leq i\leq 2N$, satisfy the pseudo-Majorana condition~(\ref{PM}) resulting in  $8N$  real components, and  $M=P_{5}$ is a real central charge.  In particular,  
since the $6$\,D tensor-multiplet is massless, $p^{\hat{\mu}}p_{\hat{\mu}}=0$,  each Kaluza-Klein mode must satisfy
\begin{equation}
p^{\mu}p_{\mu}=(\textstyle{\frac{2\pi}{R}}m)^{2}\,,
\label{mass}
\end{equation}
so that  $M$ acts as a ``mass'' operator  on $m$th Kaluza-Klein mode  with eigenvalue $\frac{2\pi}{R}m$. Massless modes, $m=0$, and massive modes, $m\neq 0$,    fit into the representations of the   little groups, $\mbox{SO}(3)\times\mbox{Sp}(N)$ and $\mbox{Spin}(4)\times\mbox{Sp}(N)\sim
\mbox{SU}(2)\times\mbox{SU}(2)\times\mbox{Sp}(N)$, separately.  
From \cite{strathdee}~(see also \cite{0004086}) the relevant representations of the  massless and massive modes are for $N=1$
\begin{equation}
\begin{array}{cl}
(2,1)\times 2^{2}~=(3,1)+(1,1)+(2,2)~~~~~~~&\mbox{:\,massless~tensor,~Maxwell,}\\
{}&{}\\
(2,1,1)\times 2^{2}=(3,1,1)+(1,1,1)+(2,1,2)~&\mbox{:\,massive~tensor}\,,
\end{array}
\end{equation}
and for $N=2$
\begin{equation}
\begin{array}{cl}
(1,1)\times 2^{4}~=(3,1)+(1,5)+(2,4)~~~~~~~&\mbox{:\,massless~tensor,~Maxwell,}\\
{}&{}\\
(1,1,1)\times 2^{4}=(3,1,1)+(1,1,5)+(2,1,4)~&\mbox{:\,massive~tensor}\,,
\end{array}
\end{equation}
where for the massless representations the tensor and Maxwell multiplets are hodge dual to each other.

%%%%%%%%%%%%%%%%%%%%%%%%%%%%%%%%%%%%%%%%%%%%%%%%%%%%%%%%%%%%%%%%%%%%%%%%%%%%%%%%%%%%%%%%%%%%%%%%%%%%%%%%%%%%%%%%%%%%%%%%%%%%%%%%%%%%%%%%%%%%%%%%%%%%%%%%%%%%%%%%%%%%%%%%%%%%%%%%%%%%%%%%%%%%%%%%%%%%%%%%%%%%%%%%%%%%%%%%%%%%%%%%%%%%%%%%%%%%%%%%%%%%%%%%%%%%%%%%%%%%%%%%%%%%%%%%%%%%%%%%%%%%%%%%%%%%%%%%%%%%%%%%%%%%%%%%%%%%%%%%%%%%%%%%%%%%%%%%%%%%%%%%%%%%%%%%%%%%%%%%%%%%%%%%%%%%%%%%%%
\section{$(1,0)$ Theory\label{sec10}}
Our two proposed five-dimensional Lagrangians for the Kaluza-Klein tower of the $6$\,D $(1,0)$ tensor-multiplet compactified on a circle are
\begin{equation}
\begin{array}{ll}
\L_{1}=\displaystyle{\sum_{m\in\Integer}}&-\textstyle{\frac{1}{4}}(F_{m\mu\nu}+i\pr mB_{m\mu\nu})(F_{-m}^{\mu\nu}-i\pr mB_{-m}^{\mu\nu}-\textstyle{\frac{1}{6}}\epsilon^{\mu\nu\lambda\rho\sigma}H_{-m\lambda\rho\sigma})\\
{}&{}\\
{}&\,+\bar{\psi}_{-mi}(i\gamma^{\mu}\partial_{\mu}+\pr m)\psi^{i}_{m}+\partial_{\mu}\phi_{m}\partial^{\mu}\phi_{-m}-(\pr m)^{2}\phi_{m}\phi_{-m}\,,
\end{array}
\end{equation}
and 
\begin{equation}
\begin{array}{ll}
\L_{2}=\displaystyle{\sum_{m\in\Integer}}&\textstyle{\frac{1}{12}}
H_{m\lambda\mu\nu}H_{-m}^{\lambda\mu\nu}-i\textstyle{\frac{\pi}{12R}}m
\epsilon^{\mu\nu\lambda\rho\sigma}B_{m\mu\nu}H_{-m\lambda\rho\sigma}\\
{}&{}\\
{}&\,+\bar{\psi}_{-mi}(i\gamma^{\mu}\partial_{\mu}+\pr m)\psi^{i}_{m}+\partial_{\mu}\phi_{m}\partial^{\mu}\phi_{-m}-(\pr m)^{2}\phi_{m}\phi_{-m}\,,
\end{array}
\end{equation}
of which  the   supersymmetry transformation rules are
\begin{equation}
\begin{array}{ccc}
\delta B_{m\mu\nu}=i\bar{\varepsilon}_{i}\gamma_{\mu\nu}\psi_{m}^{i}\,,~~~&~~~
\delta \phi_{m}=i\bar{\psi}_{mi}\varepsilon^{i}\,,~~~&~~~
\delta A_{m\mu}=i\bar{\varepsilon}_{i}\gamma_{\mu}\psi^{i}_{m}\,,\\
{}&{}&{}\\
\multicolumn{3}{c}{\delta\psi_{m}^{i}=\left\{\begin{array}{ll}
\left((\gamma^{\mu}\partial_{\mu}+i\pr m)\phi_{m}+\textstyle{\frac{1}{4}}
(F_{m\mu\nu}+i\pr mB_{m\mu\nu})\gamma^{\mu\nu}\right)\varepsilon^{i}~~~~&~~\mbox{for~~}\L_{1}\\
{}&{}\\
\left((\gamma^{\mu}\partial_{\mu}+i\pr m)\phi_{m}+\textstyle{\frac{1}{12}}
H_{m\lambda\mu\nu}\gamma^{\lambda\mu\nu}\right)\varepsilon^{i}~~~~&~~\mbox{for~~}\L_{2}
\end{array}\right. \,.}
\end{array}
\label{susy}
\end{equation}
Note that these are compatible with the reality and pseudo-Majorana conditions~(\ref{realpM}) and the invariance of the action can be shown using 
$\gamma^{\lambda\mu\nu}=\textstyle{\frac{1}{2}}
\epsilon^{\lambda\mu\nu\rho\sigma}\gamma_{\rho\sigma}$ and
\begin{equation}
\bar{\varepsilon}_{i}\gamma_{\mu_{1}}\gamma_{\mu_{2}}\cdots\gamma_{\mu_{n}}\psi^{i}_{m}=-\bar{\psi}_{mi}\gamma_{\mu_{n}}\cdots\gamma_{\mu_{2}}\gamma_{\mu_{1}}\varepsilon^{i}=-(\bar{\varepsilon}_{i}\gamma_{\mu_{1}}\gamma_{\mu_{2}}\cdots\gamma_{\mu_{n}}\psi^{i}_{-m})^{*}\,.
\end{equation}
The summation of the modes can be just over $|m|$ and $-|m|$  for any given $m\in\Integer$, as this pair alone forms an irreducible representation of the supersymmetry transformations. Hence, we may mix $\L_{1}$ and $\L_{2}$. For example,  we may replace the zero mode in $\L_{2}$ by the zero mode in $\L_{1}$ which will give the kinetic terms for both  the vector and the tensor fields.\newline

For $\L_{1}$,  the equations of motion for $B_{m\mu\nu},\,m\neq 0$ and $A_{0\mu}$ alone give  the  self-duality formula~(\ref{selfdual}) for $m\neq 0,\,m=0$ respectively.  On the other hand, $B_{0\mu\nu}$ appears only as a total derivative in $\L_{1}$ not contributing the action, and the equation of motion for $A_{m\mu},\,m\neq 0$ 
is  nothing but the divergence of the self-duality formula, and hence not a new field equation. Note also that  the zero modes correspond to  a five-dimensional super Maxwell theory.\newline

For $\L_{2}$, the equation of motion for $B_{m\mu\nu}$ is the curl of the self-duality~(\ref{curl}) and  the  gauge freedom recovers the  self-duality as we discussed in section \ref{compact}  \newline

Our two  Lagrangians, $\L_{1}$ and $\L_{2}$,  are dual to each other through the following   intermediate Lagrangian containing auxiliary two-form fields, $J_{m\mu\nu}=-J_{m\nu\mu}$,
\begin{equation}
\L_{1\leftrightarrow 2}=\displaystyle{\sum_{m\in\Integer}}\textstyle{\frac{1}{4}}J_{m\mu\nu}J^{\mu\nu}_{-m}+\textstyle{\frac{1}{2}}(F_{m\mu\nu}+i\pr mB_{m\mu\nu})(J_{-m}^{\mu\nu}+\textstyle{\frac{1}{12}}
\epsilon^{\mu\nu\lambda\rho\sigma}H_{-m\lambda\rho\sigma})\,.
\end{equation}
The equations of motion for $J_{m\mu\nu}$,  $B_{m\mu\nu}$ are  $J_{m\mu\nu}=-(F_{m\mu\nu}+i\pr mB_{m\mu\nu})$,  $J_{m\mu\nu}=-\textstyle{\frac{1}{6}}\epsilon_{\mu\nu\lambda\rho\sigma}H_{m}^{\lambda\rho\sigma}$ respectively. Depending on which expression  of the auxiliary fields  we choose to substitute,  $\L_{1\leftrightarrow 2}$ gives either  $\L_{1}$ or $\L_{2}$.  This equivalence has an analogue in three-dimensions: the free theory of Maxwell and Chern-Simons terms is equivalent to the theory of Chern-Simons and Higgs terms.

%%%%%%%%%%%%%%%%%%%%%%%%%%%%%%%%%%%%%%%%%%%%%%%%%%%%%%%%%%%%%%%%%%%%%%%%%%%%%%%%%%%%%%%%%%%%%%%%%%%%%%%%%%%%%%%%%%%%%%%%%%%%%%%%%%%%%%%%%%%%%%%%%%%%%%%%%%%%%%%%%%%%%%%%%%%%%%%%%%%%%%%%%%%%%%%%%%%%%%%%%%%%%%%%%%%%%%%%%%%%%%%%%%%%%%%%%%%%%%%%%%%%%%%%%%%%%%%%%%%%%%%%%%%%%%%%%%%%%%%%%%%%%%%%%%%%%%%%%%%%%%%%%%%%%%%%%%%%%%%%%%%%%%%%%%%%%%%%%%%%%%%%%%%%%%%%%%%%%%%%%%%%%%%%%%%%%%%%%%
\section{$(2,0)$ Theory\label{sec20}}
The $(2,0)$ tensor-multiplet contains five real scalars, $\phi^{ij},\,1\leq i,j\leq 4$, satisfying
\begin{equation}
\begin{array}{cc}
\phi^{ij}=-\phi^{ji}\,,~~~~&~~~~\phi^{ij}J_{ij}=0\,.
\end{array}
\end{equation}
With $\phi_{ij}=\phi^{kl}J_{ki}J_{lj}$ our  proposed Lagrangians  are 
\begin{equation}
\begin{array}{ll}
\L_{1}=\displaystyle{\sum_{m\in\Integer}}&-\textstyle{\frac{1}{4}}(F_{m\mu\nu}+i\pr mB_{m\mu\nu})(F_{-m}^{\mu\nu}-i\pr mB_{-m}^{\mu\nu}-\textstyle{\frac{1}{6}}\epsilon^{\mu\nu\lambda\rho\sigma}H_{-m\lambda\rho\sigma})\\
{}&{}\\
{}&\,+\bar{\psi}_{-mi}(i\gamma^{\mu}\partial_{\mu}+\pr m)\psi^{i}_{m}+\partial_{\mu}\phi^{ij}_{m}\partial^{\mu}\phi_{-mij}-(\pr m)^{2}\phi^{ij}_{m}\phi_{-mij}\,,
\end{array}
\end{equation}
and 
\begin{equation}
\begin{array}{ll}
\L_{2}=\displaystyle{\sum_{m\in\Integer}}&\textstyle{\frac{1}{12}}
H_{m\lambda\mu\nu}H_{-m}^{\lambda\mu\nu}-i\textstyle{\frac{\pi}{12R}}m
\epsilon^{\mu\nu\lambda\rho\sigma}B_{m\mu\nu}H_{-m\lambda\rho\sigma}\\
{}&{}\\
{}&\,+\bar{\psi}_{-mi}(i\gamma^{\mu}\partial_{\mu}+\pr m)\psi^{i}_{m}+\partial_{\mu}\phi^{ij}_{m}\partial^{\mu}\phi_{-mij}-(\pr m)^{2}\phi^{ij}_{m}\phi_{-mij}\,,
\end{array}
\end{equation}
and  the   supersymmetry transformation rules are
\begin{equation}
\begin{array}{cc}
\delta B_{m\mu\nu}=i\bar{\varepsilon}_{i}\gamma_{\mu\nu}\psi_{m}^{i}\,,~~~
&~~~
\delta A_{m\mu}=i\bar{\varepsilon}_{i}\gamma_{\mu}\psi^{i}_{m}\,,\\
{}&{}\\
\multicolumn{2}{l}{\delta\phi^{ij}_{m}=-i\textstyle{\frac{1}{2}}(\bar{\psi}^{i}_{m}\varepsilon^{j}-\bar{\psi}^{j}_{m}\varepsilon^{i}+\textstyle{\frac{1}{2}}J^{-1ij}\bar{\psi}_{mk}\varepsilon^{k})\,,}\\
{}&{}\\
\multicolumn{2}{c}{\delta\psi_{m}^{i}=\left\{\begin{array}{ll}
(\gamma^{\mu}\partial_{\mu}+i\pr m)\phi^{i}_{m}{}_{j}\varepsilon^{j}+
\textstyle{\frac{1}{4}}(F_{m\mu\nu}+i\pr mB_{m\mu\nu})\gamma^{\mu\nu}
\varepsilon^{i}~~~~&~~\mbox{for~~}\L_{1}\\
{}&{}\\
(\gamma^{\mu}\partial_{\mu}+i\pr m)\phi^{i}_{m}{}_{j}\varepsilon^{j}
+\textstyle{\frac{1}{12}}
H_{m\lambda\mu\nu}\gamma^{\lambda\mu\nu}\varepsilon^{i}~~~~&~~\mbox{for~~}\L_{2}
\end{array}\right. \,.}
\end{array}
\label{susy2}
\end{equation}

%%%%%%%%%%%%%%%%%%%%%%%%%%%%%%%%%%%%%%%%%%%%%%%%%%%%%%%%%%%%%%%%%%%%%%%%%%%%%%%%%%%%%%%%%%%%%%%%%%%%%%%%%%%%%%%%%%%%%%%%%%%%%%%%%%%%%%%%%%%%%%%%%%%%%%%%%%%%%%%%%%%%%%%%%%%%%%%%%%%%%%%%%%%%%%%%%%%%%%%%%%%%%%%%%%%%%%%%%%%%%%%%%%%%%%%%%%%%%%%%%%%%%%%%%%%%%%%%%%%%%%%%%%%%%%%%%%%%%%%%%%%%%%%%%%%%%%%%%%%%%%%%%%%%%%%%%%%%%%%%%%%%%%%%%%%%%%%%%%%%%%%%%%%%%%%%%%%%%%%%%%%%%%%%%%%%%%%%%%
\section{Lift to $6\,$D\label{lift}}
It is straightforward to lift our proposed actions to six-dimensions. The scalar and spinor parts are the standard ones 
\begin{equation}
\textstyle{\frac{1}{2\pi R}}\int {\rm d}^{6}x\,\,i\bar{\psi}_{i}\tilde{\gamma}^{\hat{\mu}}\partial_{\hat{\mu}}
\psi^{i}+\partial_{\hat{\mu}}\phi\partial^{\hat{\mu}}\phi\,.
\end{equation}
%%where for $(2,0)$ case, the proper indices for the scalar field,
%% $\phi^{ij}$, to be understood.
The two-form tensor part leads  for $\L_{1}$
\begin{equation}
\textstyle{\frac{1}{2\pi R}}\int {\rm d}^{6}x\,\textstyle{\frac{1}{4}}H_{5\mu\nu}H^{5\mu\nu}+\textstyle{\frac{1}{24}}\epsilon^{\mu\nu\lambda\rho\sigma}\partial_{5}B_{\mu\nu}H_{\lambda\rho\sigma}\,,
\label{B2}
\end{equation}
and for $\L_{2}$
\begin{equation}
\textstyle{\frac{1}{2\pi R}}\int {\rm d}^{6}x\,\textstyle{\frac{1}{12}}H_{\lambda\mu\nu}H^{\lambda\mu\nu}-\textstyle{\frac{1}{24}}\epsilon^{\mu\nu\lambda\rho\sigma}\partial_{5}B_{\mu\nu}H_{\lambda\rho\sigma}\,.
\label{pesc}
\end{equation}
The latter is identical to the action  by Perry and Schwarz~\cite{9611065}.  Hence, our work on $\L_{2}$ can be regarded as  
its  $(N,0),\,N=1,2$ supersymmetric extensions. The supersymmetry transformation rules  can be easily read from eqs.(\ref{susy},\ref{susy2}).  Some analysis on the canonical dimensions of the fields show $R=g_{{\rm YM}}^{2}$, where $g_{{\rm YM}}$ is the five-dimensional coupling constant~\cite{seiberg16,0004195}.\newline

Both actions in eqs.(\ref{B2},\ref{pesc}) are 
manifestly invariant under  the five-dimensional Lorentz transformations
\begin{equation}
\delta_{5{\rm D}}B_{\mu\nu}=\Lambda_{\mu}{}^{\lambda}B_{\lambda\nu}+\Lambda_{\nu}{}^{\lambda}B_{\mu\lambda}+\Lambda^{\lambda\rho}x_{\lambda}\partial_{\rho}B_{\mu\nu}\,.
\label{5Dtr}
\end{equation}  
On the other hand, it is not clear whether the actions are invariant under the rotations mixing  the sixth direction and  the other five, $\mu$, directions.   In \cite{9611065} the authors found a transformation with five-dimensional vector parameters, 
$\Lambda_{\mu}$, 
\begin{equation}
\delta B_{\mu\nu}=\textstyle{\frac{1}{6}}\Lambda{\cdot x}\epsilon_{\mu\nu\lambda\rho\sigma}H^{\lambda\rho\sigma}+ x^{5}\Lambda{\cdot\partial}B_{\mu\nu}\,,
\label{pst}
\end{equation}
which leaves the action~(\ref{pesc}) invariant upto surface terms, and it was argued that this is the remaining Lorentz symmetry so that the action possesses the full six-dimensional Lorentz symmetry.  However, in this case the transformation in eq.(\ref{pst}) lacks the usual distinction of ``spin''  and ``orbital'' parts of the Lorentz transformations as in eq.(\ref{5Dtr}).     
Furthermore, the anticommutator of the transformations reads
\begin{equation}
[\delta_{2},\delta_{1}]B_{\mu\nu}=(\Lambda_{1}{\cdot x}\Lambda_{2}^{\lambda}-\Lambda_{2}{\cdot x}\Lambda_{1}^{\lambda})(H_{\lambda\mu\nu}+\partial_{\lambda}B_{\mu\nu})\,.
\label{com21}
\end{equation}
For the transformation in eq.(\ref{pst}) to be identified with the remaining Lorentz transformations, this must be interpreted as   
the five-dimensional  Lorentz transformations~(\ref{5Dtr})  upto any possible gauge transformation.  However, direct calculation shows that this is not the case, since 
with $\Lambda_{\mu\nu}=\Lambda_{1\mu}\Lambda_{2\nu}-(1\leftrightarrow 2)$
\begin{equation}
[\delta_{2},\delta_{1}]H_{\lambda\mu\nu}=\delta_{5{\rm D}}H_{\lambda\mu\nu}+\Lambda^{\tau\kappa}x_{\tau}\partial_{\kappa}H_{\lambda\mu\nu}-3\Lambda_{\kappa [\lambda}\partial^{\kappa}B_{\mu\nu]}\,.
\end{equation}
and the right hand side can not be rescaled into\footnote{This is also impossible on shell contrary to the claim in Ref.~\cite{9611065}.}  $\delta_{5{\rm D}}H_{\lambda\mu\nu}$.  Therefore,  eq.(\ref{pst}) is  a  symmetry of the action which is not Lorentz symmetry even upto gauge transformations. Nevertheless,  the formulation by Schwarz \textit{et al.} is a certain non-covariant gauge fixing of the PST formulation~\cite{9701037,9701166,9503182,9506109}, and  the latter possesses the full $6\,\mbox{D}$ Lorentz symmetry. Furthermore it was shown that the PST action for the two-form tensor supermultiplet enjoys the six-dimensional superconformal symmetry~\cite{9711161} as well as some nontrivial local  symmetries~\cite{9701037,9710127}.  These results suggest that there is a hidden big symmetry  in the action  which combines those two symmetries and contains the transformation found by Perry and Schwarz~(\ref{pst}).  Note that  Coleman-Mandula theorem on possible symmetries of field theories applies only for massive point-like particles~\cite{PR1591251} and $6\,\mbox{D}$ two-form tensor theory is not the case, since it is massless conformal theory and the self-duality makes the distinction between the electric and the magnetic particles meaningless. \newline

%%%%%%%%%%%%%%%%%%%%%%%%%%%%%%%%%%%%%%%%%%%%%%%%%%%%%%%%%%%%%%%%%%%%%%%%%%%%%%%%%%%%%%%%%%%%%%%%%%%%%%%%%%%%%%%%%%%%%%%%%%%%%%%%%%%%%%%%%%%%%%%%%%%%%%%%%%%%%%%%%%%%%%%%%%%%%%%%%%%%%%%%%%%%%%%%%%%%%%%%%%%%%%%%%%%%%%%%%%%%%%%%%%%%%%%%%%%%%%%%%%%%%%%%%%%%%%%%%%%%%%%%%%%%%%%%%%%%%%%%%%%%%%%%%%%%%%%%%%%%%%%%%%%%%%%%%%%%%%%%%%%%%%%%%%%%%%%%%%%%%%%%%%%%%%%%%%%%%%%%%%%%%%
~\newline
~\newline
~\newline
\begin{center}
\large{\textbf{Acknowledgements}}
\end{center}
We wish to thank Dmitrij Sorokin for helpful e-mail correspondence clarifying the equivalence on PST and Schwarz \textit{et al.} formulations. K.L. is supported in part by KOSEF 1998 Interdisciplinary Research Grant 98-07-02-07-01-5.
\newline
%%%%%%%
%%\newline
%%\newline
%%%%
%%\newpage
%%%%
%%\appendix
%%\begin{center}
%%\Large{\textbf{Appendix}}
%%\end{center}
%%\setcounter{equation}{0}
%%\renewcommand{\theequation}{A.\arabic{equation}} 
%%%

\newpage
\bibliographystyle{phreport}
\bibliography{reference}

\begin{thebibliography}{10}

\bibitem{PLB27649}
{{R}. {G}{\"{u}}ven},
\newblock {Black $P$-brane Solutions of D=11 Supergravity Theory},
\newblock Phys. Lett. {\bf B276}, ~49 (1992).

\bibitem{NPB36760}
{{C}. {G}. {C}allan, {J}. {A}. {H}arvey and {A}. {S}trominger},
\newblock {Worldbrane Actions for String Solitons},
\newblock Nucl. Phys. {\bf B367}, ~60 (1991).

\bibitem{PRL713754}
{{G}. {G}ibbons and {P}. {T}ownsend},
\newblock {Vacuum Interpolation in Supergravity via Super $p$-branes},
\newblock Phys. Rev. Lett. {\bf 71}, ~3754 (1993).

\bibitem{9510053}
{{D}. {K}aplan and {J}. {M}ichelson},
\newblock {Zero Modes for the D=11 Membrane and Five-Brane},
\newblock Phys. Rev. {\bf D53}, ~3474 (1996).

\bibitem{9602071}
{{K}. {B}ecker and {M}. {B}ecker},
\newblock {Boundaries in M-Theory},
\newblock Nucl. Phys. {\bf B472}, ~221 (1996),
\newblock hep-th/9602071.

\bibitem{9610234}
{{E}. {W}itten},
\newblock {Five-Brane Effective Action in M-Theory},
\newblock J. Geom. Phys. {\bf 22}, ~103 (1997),
\newblock hep-th/9610234.

\bibitem{9611008}
{{P}. {S}. {H}owe and {E}. {S}ezgin},
\newblock {D=11 p=5},
\newblock Phys. Lett. {\bf B394}, ~62 (1997),
\newblock hep-th/9611008.

\bibitem{PRL591873}
{{R}. {F}loreanini and {R}. {J}ackiw},
\newblock {Self-Dual Fields as Charge-Density Solitons},
\newblock Phys. Rev. Lett. {\bf 59}, ~1873 (1987).

\bibitem{NPB343689}
{{B}. {M}cClain, {Y}.-{S}. {W}u and {F}. {Y}u},
\newblock {Covariant Quantization of Chiral Bosons and $\mbox{OSP}(1,1|2)$
  Symmetry},
\newblock Nucl. Phys. {\bf B343}, ~689 (1990).

\bibitem{PLB206650}
{{M}. {H}enneaux and {C}. {T}eitelboim},
\newblock {Dynamics of Chiral (Self-Dual) $p$-forms},
\newblock Phys. Lett. {\bf B206}, ~650 (1988).

\bibitem{9603031}
{{F}. {P}. {D}evecchi and {M}. {H}enneaux},
\newblock {Covariant Path Integral for Chiral P-forms},
\newblock Phys. Rev. {\bf D45}, ~1606 (1996),
\newblock hep-th/9603031.

\bibitem{9609102}
{{I}. {B}engtsson and {A}. {K}leppe},
\newblock {On Chiral P-forms},
\newblock Int. J. Mod. Phys. {\bf A12}, ~3397 (1997),
\newblock hep-th/9609102.

\bibitem{9611100}
{{P}. {P}asti, {D}. {S}orokin and {M}. {T}onin},
\newblock {On Lorentz Invariant Actions for Chiral P-Forms},
\newblock Phys. Rev. {\bf D55}, ~6292 (1997).

\bibitem{9304154}
{{J}. {H}. {S}chwarz and {A}. {S}en},
\newblock {Duality Symmetric Actions},
\newblock Nucl. Phys. {\bf B411}, ~35 (1994),
\newblock hep-th/9304154.

\bibitem{9611065}
{{M}. {P}erry and {J}. {H}. {S}chwarz},
\newblock {Interacting Chiral Gauge Fields in Six Dimensions and Born-Infeld
  Theory},
\newblock Nucl. Phys. {\bf B489}, ~47 (1997),
\newblock hep-th/9611065.

\bibitem{9701008}
{{J}. {H}. {S}chwarz},
\newblock {Coupling a Self-Dual Tensor to Gravity in Six Dimensions},
\newblock Phys. Lett. {\bf B395}, ~191 (1997),
\newblock hep-th/9701008.

\bibitem{9701037}
{{P}. {P}asti, {D}. {S}orokin and {M}. {T}onin},
\newblock {Covariant Action for a $\mbox{D}=11$ Five-Brane with the Chiral
  Field},
\newblock Phys. Lett. {\bf B398}, ~41 (1997).

\bibitem{9701166}
{{M}. {A}ganagic, {J}. {P}ark, {C}. {P}opescu and {J}. {H}. {S}chwarz},
\newblock {World Volume Action of the M Theory Five-Brane},
\newblock Nucl. Phys. {\bf B496}, ~191 (1997),
\newblock hep-th/9701166.

\bibitem{9503182}
{{P}. {P}asti, {D}. {S}orokin and {M}. {T}onin},
\newblock {Note on Manifest Lorentz and General Coordinate Invaraince in
  Duality Symmetric Models},
\newblock Phys. Lett. {\bf B352}, ~59 (1995).

\bibitem{9506109}
{{P}. {P}asti, {D}. {S}orokin and {M}. {T}onin},
\newblock {Duality Symmetric Actions with Manifest Space-Time Symmetries},
\newblock Phys. Rev. {\bf D52}, ~4277 (1995).

\bibitem{9509052}
{{P}. {P}asti, {D}. {S}orokin and {M}. {T}onin},
\newblock {Space-Time Symmetries in Duality Symmetric Models},
\newblock in {\em {Proceedings of the TMR Meeting in Leuven}}, 1995,
\newblock hep-th/9509052.

\bibitem{9701149}
{{I}. {B}andos, {K}. {L}echner, {A}. {N}urmagambetov, {P}. {P}asti, {D}.
  {S}orokin and {M}. {T}onin},
\newblock {Covariant Action for the Super-Five-Brane of M-theory},
\newblock Phys. Rev. Lett. {\bf 78}, ~4332 (1997),
\newblock hep-th/9701149.

\bibitem{9710127}
{{G}. {D}all'{A}gata, {K}. {L}echner and {M}. {T}onin},
\newblock {Covariant Actions for $N=1$,$D=6$ Supergravity Theories with Chiral
  Bosons},
\newblock Nucl. Phys. {\bf B512}, ~179 (1998).

\bibitem{9711161}
{{P}. {C}laus, {R}. {K}allosh and {A}. {V}an {P}roeyen},
\newblock {M $5$-brane and Superconformal $(0,2)$ Tensor Multiplet in $6$
  Dimensions},
\newblock Nucl. Phys. {\bf B518}, ~117 (1998).

\bibitem{9702136}
{{M}. {R}ozali},
\newblock {Matrix Theory and U-duality in Seven Dimensions},
\newblock Phys. Lett. {\bf B400}, ~260 (1997).

\bibitem{9704089}
{{M}. {B}erkooz, {M}. {R}ozali and {N}. {S}eiberg},
\newblock {Matrix Description of M-theory on $T^{4}$ and $T^{5}$},
\newblock Phys. Lett. {\bf B408}, ~105 (1997).

\bibitem{seiberg16}
N.~{S}eiberg,
\newblock {Notes on Theories with 16 Supercharges},
\newblock Proceedings of The Trieste Spring School  (1997),
\newblock {hep-th/9705117}.

\bibitem{kugotownsend}
T.~{K}ugo and P.~{T}ownsend,
\newblock {Supersymmetry and the Division Algebras},
\newblock Nucl. Phys. {\bf B221}, ~357 (1983).

\bibitem{NPB221331}
{{P}. {H}owe, {G}. {S}ierra and {P}. {T}ownsend},
\newblock {Supersymmetry in Six Dimensions},
\newblock {Nucl. Phys.} {\bf {B221}}, {~331} ({1983}).

\bibitem{strathdee}
J.~{S}trathdee,
\newblock {Extended Poincar\'{e} Supersymmetry},
\newblock Int. J. Mod. Phys. A {\bf 2}, ~273 (1987).

\bibitem{0004086}
{{C}. {H}ull},
\newblock {BPS Supermultiplets in Five Dimensions},
\newblock JHEP {\bf 0006}, 019 (2000),
\newblock hep-th/0004086.

\bibitem{0004195}
{{C}. {H}ull},
\newblock {Strongly Coupled Gravity and Duality},
\newblock hep-th/0004195.

\bibitem{PR1591251}
{{S}. {C}oleman and {J}. {M}andula},
\newblock {All Possible Symmetries of the S Matrix},
\newblock Phys. Rev. {\bf 159}, ~1251 (1967).

\end{thebibliography}
\end{document}